# CUBA: Interprocedural Context-UnBounded Analysis of Concurrent Programs (Extended Manuscript)

Peizun Liu
College of Computer and Information Science
Northeastern University
Boston, MA, United States
lpzun@ccs.neu.edu

Thomas Wahl
College of Computer and Information Science
Northeastern University
Boston, MA, United States
wahl@ccs.neu.edu

**Abstract**

A classical result by Ramalingam about synchronization-sensitive interprocedural program analysis implies that reachability for concurrent threads running recursive procedures is undecidable. A technique proposed by Qadeer and Rehof, to bound the number of context switches allowed between the threads, leads to an incomplete solution that is, however, believed to catch "most bugs" in practice. The question whether the technique can also prove the absence of bugs at least in some cases has remained largely open.

In this paper we introduce a broad verification methodology for resource-parameterized programs that observes how changes to the resource parameter affect the behavior of the program. Applied to the *context-unbounded analysis problem* (CUBA), the methodology results in partial verification techniques for procedural concurrent programs. Our solutions may not terminate, but are able to both refute and prove context-unbounded safety for concurrent recursive threads. We demonstrate the effectiveness of our method using a variety of examples, the safe of which cannot be proved safe by earlier, context-bounded methods.

*Keywords*  Interprocedural Analysis, Context Bound, Concurrent Program, Recursion, Stack

## 1  Introduction

Reasoning about systems of concurrent threads executing procedures with unbounded call stacks requires an analysis that records the state of the thread before each procedure call, in order to resume execution on the correct stack frame upon return from the call. A synchronization-sensitive analysis, on the other hand, tracks the communication between parallel threads. A landmark result by Ramalingam shows that a precise analysis respecting both procedure calls and synchronization is impossible algorithmically [37]. This implies that there is no complete reachability method for concurrent threads with shared-memory communication and recursive procedure calls.

In response to this costly (from a program analysis point of view) marriage between recursive procedures and concurrency, extensive research has been conducted on various forms of approximate solutions [6, 11, 18, 23, 27, 31, 40]. These methods sidestep the undecidability by imposing restrictions on the program behavior [6, 23], the synchronization mechanisms [11, 18], the way of switching contexts [27, 40], or the stack depth [31]. Less work exists on the alternative approach of using semi-decision procedures, for which termination may not be guaranteed. Prabhu et al. propose such a procedure to construct a proof of correctness via combining context-bounded analysis (CBA) and $k$-induction [33]. It generalizes the information obtained from CBA to construct a proof of correctness. Chaki et al. present a semi-decision procedure targeting concurrent message passing programs [13]. They tackle the undecidability using a CEGAR scheme.

Many of the preceding results exploit in one form or another the early work by Qadeer and Rehof [35], who showed that decidability of synchronization-sensitive interprocedural program analysis can be restored by imposing a *context switch bound* $k$ on the analysis: along any path, the scheduler can switch control between threads at most $k$ times. This gives rise to a principal error detection method: we algorithmically investigate the program for an increasing context bound $k$ until a bug is found. If that does not happen, empirical evidence suggests that the program may in fact be bug-free: concurrency errors tend to occur within few context switches [30, 36]. Notwithstanding such empirical evidence, it is of course highly desirable to confirm with certainty that the absence of bugs for small $k$ implies their unreachability. We call this challenge the *Context-UnBounded Analysis* problem, CUBA for short, for fixed-thread concurrent finite-state procedures communicating via shared memory.

In this paper we propose a fresh approach, the paradigm of *observation sequences*, to tackle the CUBA problem. This general paradigm applies to programs $\mathcal{P}$ that are parameterized by the amount $k$ available of some "resource", such as, in this paper, the number of permitted thread contexts along executions. Given $\mathcal{P}$ and a property $C$, the generic plan to decide whether $C$ holds in $\mathcal{P}$ *for all* values of parameter $k$ is to analyze $\mathcal{P}$ for increasing values $k = 0, 1, \ldots$. For each value, we compute an *observation* $O_k$ that we make about $\mathcal{P}$. The expectation then is that the observation sequence converges when reaching some parameter value $k_0$. If that is the case and the observations found in $O_{k_0}$ do not witness any property violation, we can conclude that $C$ holds for all $k$.

By design, observation sequences are *monotone*, which entails a number of properties useful for convergence detection. For instance, monotone sequences over a finite domain always converge. An example is a sequence that observes in $O_k$ the set of reachable program locations. In general, however, convergence is neither guaranteed, nor is it easy to detect if the sequence does converge. In particular, since observation $O_k$ is only an abstraction of the system behavior for bound $k$, the sequence may exhibit *stuttering*: its elements may form a "plateau" of the form $O_k = \ldots = O_{k+i} \subsetneq O_{k+i+1}$. The challenge for applying the observation sequence paradigm is often to distinguish sequence stuttering from convergence.

In this paper we consider observation sequences over the set $R_k$ of states reachable **under a context bound** $k$, or projections of those sets, in a constant-thread concurrent pushdown system. The sequence $(R_k)_{k=0}^{\infty}$ is naturally monotone in $k$. It is easy to see that it is also *stutter-free*: $R_{k_0} = R_{k_0+1}$ implies that the sequence in fact collapses at $k_0$ and thus converges. This immediately gives rise to a sound procedure for solving the CUBA problem: compute $R_0, R_1, \ldots$ until a plateau is observed. If no safety violation has been discovered up to this point, we can conclude that the program is safe for any context bound $k$. But of course the solution is not that simple: OS $(R_k)_{k=0}^{\infty}$ may not converge often, and computing it and detecting convergence via $R_{k_0} = R_{k_0+1}$ is expensive, as we will discuss in this paper.

An alternative partial algorithm for context-unbounded reachability analysis arises from the above note that observation sequences over a finite domain are guaranteed to converge. Since the stacks maintained by each thread can grow without bound even without any context switch, the sets $R_k$ themselves are in general infinite. We can achieve a finite domain, however, by projecting them to what we call the *visible* states of a concurrent pushdown system, which includes only the top symbol of each stack. Since the stack alphabet is finite, the domain of this projection is a finite set of tuples. The observation sequence $(\mathcal{T}(R_k))_{k=0}^{\infty}$ of visible states therefore is guaranteed to converge. Most reachability properties, including assertions inserted into a program, are formulated only over visible states and can thus be expressed over the visible state projection.

But there is no free lunch: since the CUBA problem is undecidable, it is clear that detecting convergence of the visible-state OS must be non-trivial: it will generally be hampered by stuttering. One of the technical contributions of this paper is a technique to separate stuttering from convergence. The insight is that once we have reached a plateau, i.e. a succession of equal sequence members $\mathcal{T}(R_k) = \ldots = \mathcal{T}(R_{k+i})$, the *first* (if any) visible state that shows up later in the sequence must have a special form: it must be triggered by particular stack operations. We call such visible states *generators*. Due to their special form, the reachability of new generator states can in many cases be ruled out, given the current information provided by $\mathcal{T}(R_{k+i})$. The theorem then is that, once we reach a plateau *and* we can rule out the reachability of new generator states, the sequence has converged and the CUBA problem is solved.

Towards practical implementation of these techniques to solve the CUBA problem, we face another difficulty: since the sets $R_k$ can be infinite, they must be represented symbolically. The data structure proposed by Bouajjani et al. [8] for this purpose and later used in [35, 38] is the pushdown store automaton. These automata were shown to have worst-case size exponential in $k$. This can make their construction expensive in practice, even if we later project each $R_k$ to a finite set such as $\mathcal{T}(R_k)$. We thus have to find alternative ways of maintaining these sets.

To this end we observe that, while the set $\bigcup_{k=1}^{\infty} R_k$ of all reachable global states, an infinite union, is often infinite, the sets $R_k$ of states reachable with context bound $k$ may well be finite. We call this condition *finite context reachability* (FCR). Since our approach based on observation sequences is iterative and always maintains only a set of global states reachable under the current context bound $k$, FCR allows us to represent sets of states using compact data structures for finite sets, such as BDDs or even extensional lists or sets. This facilitates implementing the technique, and makes operations such as equivalence checks vastly more efficient than using pushdown store automata.

The question that remains is: how do we decide whether FCR holds for a given program $\mathcal{P}$? We show in this paper that, for a sequential pushdown system, if the set of states reachable *from all stacks of size $\leq 1$* is finite, then the set of states reachable *from any **one** state with arbitrary stack size* is finite. By induction, then, if this condition holds for all threads' pushdown systems, all sets $R_k$ are finite.

We conclude this paper with a number of experiments on concurrent procedural programs, most featuring unbounded recursion, many used in previous related work. We demonstrate on these programs the effectiveness and relative efficiency of our proposed techniques. The main insights from our experiments are: our benchmark programs (i) exhibit fairly small context bounds $k_0$ at which the visible-states observation sequence $(\mathcal{T}(R_k))_{k=0}^{\infty}$ converges, thus allowing us to prove their safety, and in fact in about the same or less time than previous context-*bounded* methods; and (ii) almost all satisfy the FCR condition, which facilitates the analysis.

In summary, we present in this paper a partial technique to decide reachability in concurrent procedural programs with possibly unbounded recursion, a generally undecidable problem. While our technique may not terminate, in contrast to context-bounded methods it can both refute *and* prove safety properties. The technique is based on the simple yet elegant paradigm of observation sequences, which we introduce in Sec. 3. By their nature, techniques based on observation sequences find the smallest value $k$ for which an error occurs or the sequence converges, as the case may be.

# 2 Program Model and Problem Definition

CUBA is intended for concurrent pushdown systems, an abstract finite-state model for multi-threaded Boolean programs, which in turn result from predicate abstractions of source code. App. B describes a simplified language for Boolean programs, as used in some examples in this paper. In this section we define concurrent pushdown systems, and recall the basics of context-bounded reachability analysis.

## 2.1 Pushdown Systems

A (sequential) *pushdown system (PDS)* is a tuple $(Q, \Sigma, \Delta, q^I)$ consisting of the set of *shared states* $Q$, the *stack alphabet* $\Sigma$, the *pushdown program* $\Delta$, and the *initial shared state* $q^I \in Q$. Set $\Sigma$ does not contain the empty word symbol $\varepsilon$. Let $\Sigma^{\leq 1} = \{w \in \Sigma^* : |w| \leq 1\} = \Sigma \cup \{\varepsilon\}$ and $\Sigma^{\leq 2} = \{w \in \Sigma^* : |w| \leq 2\}$. The set $\Delta$ of *actions* is a subset of $(Q \times \Sigma^{\leq 1}) \times (Q \times \Sigma^{\leq 2})$. We write $(q, w) \to (q', w')$ for $((q, w), (q', w')) \in \Delta$.

**Semantics.** A *state* of a PDS is an element of $Q \times \Sigma^*$, written in angle brackets: $\langle q \,|\, w \rangle$; state $c^I = \langle q^I \,|\, \varepsilon \rangle$ is *initial*. Actions cause states to change; we extend relation $\to$ to states to capture *steps* of the PDS:

(a) Given a state $s = \langle q \,|\, \sigma_1..\sigma_z \rangle$, i.e. with a non-empty stack of size $z \geq 1$ with $\sigma_1$ the "top", we have $s \to s'$, where the successor state $s'$ depends on the action as follows:
   **action** $(q, \sigma_1) \to (q', \varepsilon)$:
   if $z = 1$, then $s' = \langle q' \,|\, \varepsilon \rangle$, else $s' = \langle q' \,|\, \sigma_2..\sigma_z \rangle$. This action **pops** $\sigma_1$ off the stack (modeling e.g. a terminating procedure);
   **action** $(q, \sigma_1) \to (q', \sigma')$ such that $\sigma' \in \Sigma$:
   $s' = \langle q' \,|\, \sigma' \sigma_2..\sigma_z \rangle$. This action **overwrites** $\sigma_1$ by $\sigma'$ (modeling e.g. a local variable change in the currently executing procedure);
   **action** $(q, \sigma_1) \to (q', \rho_0 \rho_1)$ such that $\rho_0, \rho_1 \in \Sigma$:
   $s' = \langle q' \,|\, \rho_0 \rho_1 \sigma_2..\sigma_z \rangle$. This action **overwrites** $\sigma_1$ by $\rho_1$ and **pushes** $\rho_0$ on the stack (modeling e.g. a procedure call; the callee changes its pc).
   No other action is enabled in state $s$.

(b) Given a state $s = \langle q \,|\, \varepsilon \rangle$, i.e. with the empty stack, the only enabled actions are of the form $(q, \varepsilon) \to (q', w')$ for $w' \in \Sigma^{\leq 1}$, changing $s$ to $s' = \langle q' \,|\, w' \rangle$. These actions are either overwrites that only change the shared state (when $w' = \varepsilon$), or they are pushes (when $|w'| = 1$).

As can be seen, a stack is initially empty. When using stacks to model procedure calls, there is typically a main thread that creates worker threads and passes to them the name of the function to be executed. In examples, we mostly omit the main thread (it is irrelevant for our purposes) and directly start each stack to contain a single symbol (interpreted as the name of the passed function). The stack will then normally not be empty until the program terminates, while our model is more general and allows stacks to become intermittently empty. Finally, our rules for push and pop actions allow the shared state to change.

**Reachability in PDS.** Let $\to^*$ denote the reflexive transitive closure of $\to$. State $s$ of a PDS is *reachable* if $c^I \to^* s$. The reachability problem for PDS is decidable [16, 38]; the (possibly infinite) set of reachable states of $\mathcal{P}$ can be symbolically represented using *pushdown store automata* (PSA) [38]. App. C presents more details about PSA.

## 2.2 Concurrent Pushdown Systems

A *concurrent pushdown system* (CPDS) is a fixed-thread, asynchronous combination of sequential PDS. Formally, given $n \in \mathbb{N}$, a CPDS $\mathcal{P}^n$ is a collection of $n$ PDS $\mathcal{P}_i = (Q, \Sigma_i, \Delta_i, q^I)$, $1 \leq i \leq n$. The set $Q$ of shared states is the same for all $\mathcal{P}_i$, as is the initial shared state $q^I$. The PDS $\mathcal{P}_i$ have individual stack alphabets and pushdown programs. A *state* of a CPDS is an element of $Q \times \Sigma_1^* \times \ldots \times \Sigma_n^*$, written in angle brackets form $\langle q \,|\, w_1, \ldots, w_n \rangle$; this indicates that $q$ is the shared state, and thread $i \in \{1, \ldots, n\}$ has stack contents $w_i \in \Sigma_i^*$. State $\langle q^I \,|\, \varepsilon, \ldots, \varepsilon \rangle$ is *initial*.

Given a state $s = \langle q \,|\, w_1, \ldots, w_n \rangle$, we refer to $(q, w_i)$ as thread $i$'s *thread state*. In addition, of interest is frequently the *thread-visible state*, which comprises the part of $s$ that is visible to a thread. For example, the executability of a thread's actions depends only on its visible state. Formally, let $\mathcal{T}: \Sigma^* \to \Sigma^{\leq 1}$ be the function

$$\mathcal{T}(w) = \begin{cases} \sigma_1 & \text{if } w = \sigma_1..\sigma_z \\ \varepsilon & \text{otherwise } (w = \varepsilon) \end{cases} . \quad (1)$$

which extracts the top symbol, if any, from a stack. We extend $\mathcal{T}$ to act on a thread state via $\mathcal{T}(q, w) = (q, \mathcal{T}(w))$, and on state $s$ via $\mathcal{T}(s) = \langle q \,|\, \mathcal{T}(w_1), \ldots, \mathcal{T}(w_n) \rangle$; the latter is called $s$'s *visible state (projection)*.

A *step* of the CPDS $\mathcal{P}^n$ is an action performed by one of the $\mathcal{P}_i$, which changes $\mathcal{P}^n$'s current state $s = \langle q \,|\, w_1, \ldots, w_n \rangle$, as follows. First, $i \in \{1, \ldots, n\}$ is nondeterministically chosen. Second, an action in $\Delta_i$ is nondeterministically chosen. Third, the action is applied to state $\langle q \,|\, w_i \rangle$ of PDS $\mathcal{P}_i$ (if enabled; otherwise the CPDS step is a no-op). This results in a new state $\langle q' \,|\, w_i' \rangle$ of $\mathcal{P}_i$. Fourth, the new CPDS state $s'$ is obtained from $s$ by replacing $q$ by $q'$ and $w_i$ by $w_i'$. We say the CPDS makes a step from $s$ to $s'$ *triggered* by thread $i$.

The above step semantics describes a binary relation $\to$ on the set of states. Let again $\to^*$ be the reflexive transitive closure. State $s$ of CPDS $\mathcal{P}^n$ is *reachable* if $\langle q^I \,|\, \varepsilon, \ldots, \varepsilon \rangle \to^* s$. We denote by $R$ the (often infinite) set of states reachable in $\mathcal{P}^n$. The reachability problem for concurrent pushdown systems is undecidable [37].

## 2.3 Context-(Un)Bounded Reachability

Consider a *path* generated by a CPDS, i.e. a sequence of states pairwise related by $\to$. Each step along the path is triggered by exactly one thread. A *context* is a sequence of consecutive steps along a path that are all triggered by the same thread.

For $k \in \mathbb{N}$ we say state $s$ is *reachable with context bound $k$* (or *within $k$ contexts*, for short) if there exists a path from $\langle q^I \mid \varepsilon, \ldots, \varepsilon \rangle$ to $s$ that consists of at most $k$ contexts.[1] Context-bounded reachability analysis—the problem of determining, given a CPDS $\mathcal{P}^n$, a state $s$ and a number $k$, whether $s$ is reachable in $\mathcal{P}^n$ within $k$ contexts—is practically interesting in part because it is decidable [35].

In contrast, in this paper we are concerned with partial solutions to the undecidable CONTEXT-UNBOUNDED (reachability) ANALYSIS problem, CUBA, for CPDS. In the sequel, resource bound $k$ will therefore always play the role of the number of contexts. We denote by $R_k$ the set of states reachable within $k$ contexts. Despite the bound, set $R_k$ can still be infinite: $k$ does not restrict the number of steps a single thread can trigger within a context.

## 3 Analyzing Programs using Observation Sequences

The vehicle for our approach to context-unbounded concurrent reachability analysis is the simple paradigm of *observation sequences* (OS), whose intuition was introduced in Sec. 1. In this section we summarize the key technical points, mostly independently of the application to the CUBA problem.

The paradigm considers program( model)s $\mathcal{P}$ that are parameterized by the amount $k$ available of some *resource*. Resource bounds may be physical or logical, and they may be imposed only for analysis purposes (such as in this paper). Let $C$ be a property of interest, which we wish to establish in $\mathcal{P}$ for all values of $k$. The generic plan is to analyze $\mathcal{P}$ for increasing values $k = 0, 1, \ldots$ and to compute, for each $k$, an *observation* $O_k$ about $\mathcal{P}$, such as what states are reachable under bound $k$.

**Definition 1.** *An **observation sequence** is a sequence $(O_k)_{k=0}^\infty$ such that the following properties hold:*

- ***monotonicity:*** *for all $k$, $O_k \subseteq O_{k+1}$.*
- ***computability:*** *for all $k$, $O_k$ is computable.*
- ***expressibility:*** *property $C$ is **expressible** over $(O_k)_{k=0}^\infty$, which means that, for all $k$, $O_k$ contains enough information to decide whether $C$ holds for $\mathcal{P}$ under bound $k$, and this question is algorithmically decidable.*

*Sequence $(O_k)_{k=0}^\infty$ has **domain** $D$ if $\cup_{k=0}^\infty O_k \subseteq D$.*

**Example 2.** *Consider the two-thread CPDS in Fig. 1 (left). If $k$ denotes the maximum number of thread contexts allowed per execution path, then let $R_k$ be the set of global states reachable up to bound $k$, and $\mathcal{T}(R_k)$ be the set of **visible** states reachable up to bound $k$, i.e. the projections of states in $R_k$ via function $\mathcal{T}$. Both $(R_k)_{k=0}^\infty$ and $(\mathcal{T}(R_k))_{k=0}^\infty$ are observation sequences: they are monotone by construction, and computable by [35]. Any reachability property such as safety up to context bound $k$ is expressible over $(R_k)_{k=0}^\infty$, while $(\mathcal{T}(R_k))_{k=0}^\infty$ is more restricted but still permits context-bounded properties stating shared-state reachability, or mutually exclusive local-state reachability.*

Table 1 defines the key concepts that are frequently used in this paper. In addition, an OS *diverges* if it does not converge. This and all terms defined in Table 1 are with respect to an input program over which the OS is defined.

***Observation Sequences: Basic Properties.*** The phenomena of plateaus, stuttering, collapse and convergence of observation sequences interact in several interesting ways, summarized in this section in basic properties. Proofs of these properties are included in App. A.

**Property 3.** *An OS $(O_k)_{k=0}^\infty$ over a finite domain converges.*

Convergence alone does not imply that the limit of sequence $(O_k)_{k=0}^\infty$ can be computed, due to stuttering. In the absence of stuttering, however, reaching a plateau is tantamount to convergence:

**Property 4.** *If OS $(O_k)_{k=0}^\infty$ does not stutter at $k_0$ and plateaus at $k_0$, then it collapses at $k_0$.*

**Example 5.** *Consider Fig. 1 again. Sequence $(\mathcal{T}(R_k))_{k=0}^\infty$ has a finite domain and thus converges. As the right-hand side of the figure shows, it plateaus at $k = 2$ ($\mathcal{T}(R_2) = \mathcal{T}(R_3)$) and $k = 5$; the goal in this paper will be to show that it in fact collapses at $k = 5$. In contrast, as we shall see in Sec. 4, sequence $(R_k)_{k=0}^\infty$ is stutter-free but diverges.*

---

**Scheme 1** Verifying $C$ for unbounded $k$ using some observation sequence $(O_k)_{k=0}^\infty$

1: **for** $k := 1$ **to** $\infty$ **do**
2:    **if** $O_k$ witnesses a violation of $C$ under bound $k$ **then**
3:       **return** "error reachable with resource amount $k$"
4:    **if** $O_{k-1} = O_k$ **then**
5:       **return** "$C$ holds (for any resource amount)"

---

Consider now the algorithmic Scheme 1 for verifying a property $C$ using an OS $(O_k)_{k=0}^\infty$. The scheme does the obvious: it increases $k$ (with $O_0$ suitably initialized) until the sequence seems to have converged, checking for errors on the way. We often refer to the iterations of the main loop in Line 1 as *rounds* of Scheme 1. Whether this scheme is implementable, and what it is good for, depends on features of the observation sequence:

(a) Since $C$ is expressible over $(O_k)_{k=0}^\infty$ (Def. 1), Line 2 is computable.
(b) If equality in the sequence domain is decidable, then so is the test in Line 4. (This condition is not trivial if the domain of $(O_k)_{k=0}^\infty$ is infinite.)
(c) If $C$ is violated (i.e., for some $k$), then Scheme 1 terminates (witnessed by $O_k$).

---

[1] An alternative definition uses the notion of a context *switch* bound [35]. We chose the equivalent but simpler formulation via context bounds. The two are related in the obvious way.

$$
\begin{aligned}
Q &= \{\, 0, 1, 2, 3 \,\} \\
\Sigma_1 &= \{\, 1, 2 \,\} \\
\Sigma_2 &= \{\, 4, 5, 6 \,\} \\
\Delta_1 &= \{\, f_1 : (0,1) \to (1,2), \\
         &\qquad f_2 : (3,2) \to (0,1) \,\} \\
\Delta_2 &= \{\, b_1 : (0,4) \to (0,\varepsilon), \\
         &\qquad b_2 : (1,4) \to (2,5), \\
         &\qquad b_3 : (2,5) \to (3,46) \,\} \\
q^I &= 0
\end{aligned}
$$

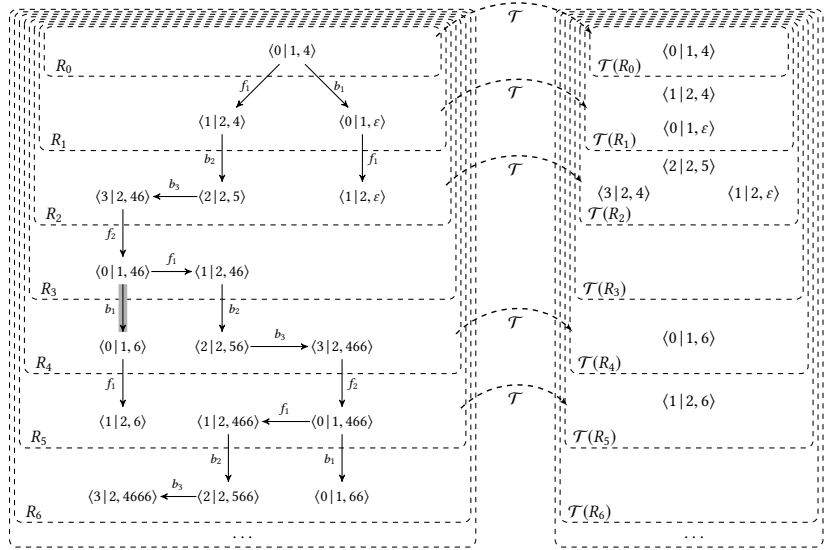

**Figure 1.** A two-thread CPDS $\mathcal{P}^2$ (left) and its *reachability table* (right). We have $\mathcal{P}^2 = \{\mathcal{P}_1, \mathcal{P}_2\}$ with $\mathcal{P}_i = (Q, \Sigma_i, \Delta_i, q^I)$ for $i = 1, 2$; the initial state is $\langle 0 \,|\, 1, 4 \rangle$. The table on the right shows the sets $R_k \setminus R_{k-1}$ and $\mathcal{T}(R_k) \setminus \mathcal{T}(R_{k-1})$ of reachable states and of reachable visible states, resp., that are new at bound $k$, for $k = 1, \ldots, 6$.

**Table 1.** Basic terminology concerning an observation sequence $(O_k)_{k=0}^{\infty}$

| Terminology: $(O_k)_{k=0}^{\infty}$ ... | Definition | Comments |
| --- | --- | --- |
| ... *plateaus* at $k_0$ | $O_{k_0} = O_{k_0+1}$ | pauses *or* stops growing |
| ... *stutters* at $k_0$ | $O_{k_0} = O_{k_0+1} \wedge \exists_{k > k_0}\, O_k \subsetneq O_{k+1}$ | pauses but doesn't stop growing |
| ... *collapses* at $k_0$ | $\forall_{k \geq k_0}\, O_{k_0} = O_k$ | stops growing |
| ... *converges* | $\exists_{k_0}\, (O_k)_{k=0}^{\infty}$ collapses at $k_0$ | collapses at some $k_0$ |
| ... *is stutter-free* | $\forall_{k_0}\, \neg(\,(O_k)_{k=0}^{\infty}$ stutters at $k_0\,)$ | does not stutter at any $k_0$ |

(d) If $(O_k)_{k=0}^{\infty}$ converges, then Scheme 1 terminates as well, since then eventually $O_{k-1} = O_k$.

(e) If $(O_k)_{k=0}^{\infty}$ is stutter-free, then the output in Line 5 is correct.

We summarize the special case that *all* these preconditions are met, as follows:

**Property 6.** *If OS $(O_k)_{k=0}^{\infty}$ converges and is stutter-free, and equality in the domain of $(O_k)_{k=0}^{\infty}$ is decidable, then any property expressible over $(O_k)_{k=0}^{\infty}$ is decidable.*

We take a systematic look at the interaction between stuttering and convergence. The adjacent

| Stutter-free: | Converges: | |
| --- | --- | --- |
| | yes | no |
| yes | ✗ | $(R_k)_{k=0}^{\infty}$ |
| no | $(\mathcal{T}(R_k))_{k=0}^{\infty}$ | – |

table shows the four combinations possible between these two features, and where we "find" them in the case that $k$ denotes a context bound for concurrent threads running recursive procedures. The most powerful observation sequences do not stutter and are guaranteed to converge. By Prop. 6 and the undecidability of the CUBA problem, such an OS **does not exist** for this problem (indicated by "✗" in the table). A stutter-free OS that is not guaranteed to converge gives rise to a possibly non-terminating but partially correct algorithm derived from Scheme 1, for instance the sequence $(R_k)_{k=0}^{\infty}$ in Ex. 2. A stuttering OS that always converges, such as the sequence $(\mathcal{T}(R_k))_{k=0}^{\infty}$, is algorithmically more difficult to use. Finally, OS that may not converge *and* suffer from stuttering are least useful and not considered in this paper.

## 4 CUBA using Observation Sequences

To illustrate the use of observation sequences for context-unbounded analysis, we begin with the simple instance of sequence $(R_k)_{k=0}^{\infty}$, which maps the context parameter $k$ to the full set $R_k$ of states reachable under that context bound (see also Ex. 2). $(R_k)_{k=0}^{\infty}$ does not suffer from plateaus that "fake convergence", as is straightforward to show (see App. A):

**Lemma 7.** $(R_k)_{k=0}^{\infty}$ *is stutter-free: for all $k_0$, if $R_{k_0} = R_{k_0+1}$, then for all $k \geq k_0$, $R_k = R_{k+1}$.*

By the discussion from Sec. 3, Scheme 1 instantiated with the stutter-free OS $(R_k)_{k=0}^{\infty}$, denoted *Scheme 1*$(R_k)$, is a partially correct algorithm to decide reachability in concurrent pushdown systems for unbounded numbers of thread contexts. Consider

**Example 8.** *The two-thread program shown in Fig. 2 is taken, with slight adaptations, from [33]; the approach presented there fails to terminate on that example. The example is small but non-trivial in that both threads have stacks, and each can grow without bound, even without any context switch. Scheme 1 ($R_k$) does not terminate for $k = 2$: $R_1 \subsetneq R_2$. Indeed, consider $c = \langle 1 \,|\, 4, 9 \rangle$, where $x = 1$, thread* foo *is spinning in the* **while** *loop, and* bar *has reached the end of its program. This state is reachable with two contexts (belongs to $R_2$), as witnessed by the path*

$$\langle \bot \,|\, 2, 6 \rangle \xrightarrow{f_1} \langle 1 \,|\, 2, 6 \rangle \xrightarrow{f_{2b}} \langle 1 \,|\, 4, 6 \rangle \xrightarrow{b_{6b}} \langle 1 \,|\, 4, 8 \rangle \xrightarrow{b_{8b}} \langle 1 \,|\, 4, 9 \rangle \,.$$

*State $c$ is obviously not reachable with one context (does not belong to $R_1$), since both threads have left their initial state. However, Scheme 1 ($R_k$) does terminate for $k = 3$: $R_2 = R_3$.*

```
1  decl x;

   void foo() {
2    if (*)
3      call foo();
4    while (x) {}
5    x := 1;
   }

   void bar() {
6    if (*)
7      call bar();
8    while (!x) {}
9    x := 0;
   }

   void main() {
10   thread_create(&foo);
11   thread_create(&bar);
   }
```

$\Delta_1 = \{$
$f_0 : (\bot, 2) \to (x, 2)$
$f_{2a} : (x, 2) \to (x, 3)$
$f_{2b} : (x, 2) \to (x, 4)$
$f_3 : (x, 3) \to (x, 24)$
$f_{4a} : (1, 4) \to (1, 4)$
$f_{4b} : (0, 4) \to (0, 5)$
$f_5 : (x, 5) \to (1, \varepsilon) \}$

$\Delta_2 = \{$
$b_0 : (\bot, 6) \to (x, 6)$
$b_{6a} : (x, 6) \to (x, 7)$
$b_{6b} : (x, 6) \to (x, 8)$
$b_7 : (x, 7) \to (x, 68)$
$b_{8a} : (0, 8) \to (0, 8)$
$b_{8b} : (1, 8) \to (1, 9)$
$b_9 : (x, 9) \to (0, \varepsilon) \}$

**Figure 2.** A CPDS consisting of two procedures foo and bar, and shared Boolean variable x, initialized nondeterministically. We formalize this program as $\mathcal{P}^2 = \{\mathcal{P}_1, \mathcal{P}_2\}$ (procedure main omitted) with $\mathcal{P}_i = (Q, \Sigma_i, \Delta_i, q^I)$ for $i = 1, 2$, $Q = \{\bot, 0, 1\}$, $\Sigma_1 = \{2, 3, 4, 5\}$, $\Sigma_2 = \{6, 7, 8, 9\}$. Shared state $\bot$ models the initial nondeterminism in x: $q^I = \langle \bot \,|\, 2, 6 \rangle$. Actions in the figure that mention symbol $x$ exist for $x = 0$ and for $x = 1$.

In practice, the utility of Scheme 1 ($R_k$) is limited by two factors. The first is that the efficiency of implementing it suffers from the complexity of computing and representing the sets $R_k$: once computed, they are used in Line 4 in a comparison. If we use pushdown store automata $A_k$ to represent the $R_k$, the comparison requires an automaton equivalence check, which can be done via language containment: $L(A_k) \cap \overline{L(A_{k-1})} = \emptyset$. Efficient complementation requires deterministic automata. Since the NFAs $A_k$ already have worst-case size exponential in $k$, the determinization can potentially yield a doubly-exponential running time. These estimates suggest that, using pushdown store automata, Scheme 1 ($R_k$) cannot be expected to scale to large pushdown systems. We address this problem in Sec. 5.

The second, more fundamental limitation is that $(R_k)_{k=0}^{\infty}$ cannot eventually hit a plateau for every input program: due to undecidability, there must be instances of divergence. We address this problem in the rest of the present section.

### 4.1 CUBA using Visible State Reachability

Since we cannot have both stutter-freeness and a convergence guarantee for an observation sequence that solves the CUBA problem (without restricting the input language), we investigate in this section a sequence with features diagonally opposite to those of $(R_k)_{k=0}^{\infty}$: guaranteed convergence but potential stuttering. By Prop. 3, convergence is guaranteed by a finite domain. To this end, let $(\mathcal{T}(R_k))_{k=0}^{\infty}$ be the sequence that maps the context parameter $k$ to the *finite* set $\mathcal{T}(R_k)$ of *visible* states (Sec. 2.2) reachable under that context bound (see also Ex. 2). Set $\mathcal{T}(R_k)$ is computed by projecting the full set $R_k$ to the tuple of shared state and the top of the stacks of all threads:

$$\mathcal{T}(R_k) = \{\langle q \,|\, \mathcal{T}(w_1), \ldots, \mathcal{T}(w_n) \rangle : \langle q \,|\, w_1, \ldots, w_n \rangle \in R_k \} \,.$$

Guaranteeing convergence, this sequence must exhibit stuttering behavior, for some programs. We give an example of this phenomenon.

**Example 9.** *We revisit the example shown in Fig. 1. The sequence $(\mathcal{T}(R_k))_{k=0}^{\infty}$ plateaus at $k = 2$: $\mathcal{T}(R_2) = \mathcal{T}(R_3)$. In the next step we find out that this plateau is merely due to stuttering: $\mathcal{T}(R_3) \subsetneq \mathcal{T}(R_4)$. Another plateau emerges at $k = 5$. The table shows no reachable visible states beyond $\mathcal{T}(R_5)$.*

In this section we design a technique that easily proves that $(\mathcal{T}(R_k))_{k=0}^{\infty}$ collapses at $k = 5$, for the program in Fig. 1. The problem we have to solve is to distinguish plateaus that merely signal stuttering from those that signal sequence convergence. We will accomplish this by replacing the test $\mathcal{T}(R_{k-1}) = \mathcal{T}(R_k)$ in Line 4 of Scheme 1 (which would lead to incorrect answers) by a stronger one that rules out stuttering.

#### 4.1.1 Stuttering detection using generators

The idea for designing a sufficient condition for the absence of stuttering at $k$ is as follows. Assume, by contraposition, that $\mathcal{T}(R_{k-1}) = \mathcal{T}(R_k) \subsetneq \mathcal{T}(R)$. It is clear that, once any $g \in \mathcal{T}(R) \setminus \mathcal{T}(R_k)$ has been encountered (for some larger $k$), its discovery may generate many more heretofore unseen states that follow in the wake of $g$. But what about the *first* time a new state in $\mathcal{T}(R) \setminus \mathcal{T}(R_k)$ is encountered after a plateau? Intuition suggests that such a state might be of a special form. If we can rule out the reachability of any new states of this form, for any larger $k$, then the same holds for all those *first* new encountered states and hence for *any* new states: the sequence has converged.

Before we put this intuition to the test, let us formalize the idea. We slightly generalize the concept of generators to any set $\mathcal{G}$ of visible states with the following property: if, upon encountering a plateau, all reachable visible states in $\mathcal{G}$ have been reached, the sequence converges. More precisely:

**Definition 10.** *A set $\mathcal{G}$ of visible states is a **generator set** if the following condition holds for every $k$: if $\mathcal{T}(R_{k-1}) = \mathcal{T}(R_k)$ and $\mathcal{G} \cap \mathcal{T}(R) \subseteq \mathcal{T}(R_k)$, then $\mathcal{T}(R_k) = \mathcal{T}(R)$.*

It is immediate from the definition that the notion of being a generator set is *upward closed*: any superset of a generator set is also a generator set. Thus, there is a natural desire to keep generator sets minimal.

Def. 10 suggests the following strategy for stuttering detection using generators:

(a) Statically define a set $\mathcal{G}$ of visible states, and prove that it is a generator set.
(b) During the execution of Scheme 1, if the sequence plateaus at $k - 1$, prove that all reachable generators (the elements of $\mathcal{G} \cap \mathcal{T}(R)$) have been reached. If so, terminate with $\mathcal{T}(R_k)$ as reachable visible states.

So far the discussion in Sec. 4.1 has been very generic. Step (a) above of course depends on the application and is the topic of Sec. 4.1.2. Step (b) seems preposterous: it involves the set $\mathcal{T}(R)$ of reachable visible states whose determination is the very goal of this section—are we caught in a cyclic argument? We will address this step in Sec. 4.1.3 and 4.1.4.

### 4.1.2 A generator set for CUBA

Given an application domain for the observation sequences paradigm, finding a set $\mathcal{G}$ with the strong guarantees suggested by Def. 10 requires good intuition about that domain. For the case of concurrent pushdown systems, we define $\mathcal{G}$ to be the visible states $\langle q | \sigma_1, \ldots, \sigma_n \rangle$ in which at least one thread-visible state $(q, \sigma_i)$ may have emerged as the result of a **pop** action, in the following sense:

$$\mathcal{G} = \{\langle q | \sigma_1, \ldots, \sigma_n \rangle : \text{ there exists } i \text{ s.t.} \\ (q, \varepsilon) \text{ is the target of a } pop \text{ edge in } \Delta_i \text{ \textbf{and}} \\ (\sigma_i = \varepsilon \text{ \textbf{or} } (?, ?\sigma_i) \text{ is the target of a } push \text{ edge in } \Delta_i)\}, \quad (2)$$

where ? stands for arbitrary shared states or stack symbols, resp. Given $i$ as in (2), visible thread states $(q, \sigma_i)$ are those that can emerge from a pop: the shared state is the target of a pop, **and** the resulting stack is either empty **or** has a top symbol that was overwritten as part of an earlier push. These are the symbols $\rho_1$ in a push action $(q, \sigma) \to (q', \rho_0 \rho_1)$.

Note that $\mathcal{G}$ is defined purely syntactically, via the pushdown programs. As a consequence, stack symbols $\sigma_j$ for threads $j \neq i$ can be arbitrary and in particular unreachable—Eq. (2) does not restrict them in any way. This problem is solved as part of step (b) mentioned above, where we aim to project set $\mathcal{G}$ to its reachable fragment; see Sec. 4.1.3.

**Theorem 11.** *Set $\mathcal{G}$ defined in (2) is a generator set.*

**Proof**: The condition in Def. 10 has the propositional form $X \wedge Y \Rightarrow Z$, which is equivalent to $X \wedge \neg Z \Rightarrow \neg Y$; we prove the latter form. Let therefore $k$ be such that $\mathcal{T}(R_{k-1}) = \mathcal{T}(R_k)$; further $\mathcal{T}(R_k) \neq \mathcal{T}(R)$, i.e. $\mathcal{T}(R) \setminus \mathcal{T}(R_k) \neq \emptyset$; we prove $\mathcal{G} \cap \mathcal{T}(R) \nsubseteq \mathcal{T}(R_k)$, i.e. $\mathcal{G} \cap (\mathcal{T}(R) \setminus \mathcal{T}(R_k)) \neq \emptyset$.

Sequence $(\mathcal{T}(R_k))_{k=0}^{\infty}$ stutters at $k - 1$; due to guaranteed convergence, the plateau eventually comes to an end (but could be long). Let $K$ mark the "last index of equality":

$$K = \max\{k' \in \mathbb{N} : k' \geq k \wedge \mathcal{T}(R_{k'-1}) = \mathcal{T}(R_{k'})\} \, .$$

So we have $\mathcal{T}(R_{k-1}) = \mathcal{T}(R_k) = \ldots = \mathcal{T}(R_K) \subsetneq \mathcal{T}(R_{K+1})$. Let $t$ be a state selected from $R_{K+1}$ such that $\mathcal{T}(t) \notin \mathcal{T}(R_K)$, as follows. Along the path $p$ to $t$, let $i$ be the identity of the thread executing in context $K + 1$, and $a$ be the final action executed by $i$ as it reaches $t$. In the following drawing, $\circ$ represents context switches, and $\to$ denotes steps (global transitions), as in Sec. 2.2:

$$p : \quad \cdots \circ \underbrace{\to \cdots \to}_{K-1} \circ \underbrace{\to \cdots \to}_{K} \circ \underbrace{\xrightarrow{i} \cdots \xrightarrow{i} s \xrightarrow{i: a} t}_{K+1}$$

We select $t$ such that $\mathcal{T}(t)$ is the *first* new visible state along the segment of $p$ inside context $K + 1$. This means that, for the state $s$ preceding $t$ along $p$, $\mathcal{T}(s) \in \mathcal{T}(R_K) = \mathcal{T}(R_{K-1})$. Hence, there exists a state $s' \in R_{K-1}$ such that $\mathcal{T}(s') = \mathcal{T}(s)$.

We now show that $\mathcal{T}(t) \in \mathcal{G}$, which proves $\mathcal{G} \cap (\mathcal{T}(R) \setminus \mathcal{T}(R_k)) \neq \emptyset$. To this end, we distinguish action types for $a$: if $a$ is a *push* or an *overwrite*, then firing $a$ from $\mathcal{T}(s')$ instead of $\mathcal{T}(s)$ yields the same successor visible state, namely $\mathcal{T}(t)$ (the successor does not depend on the stack history in these cases). This is a contradiction, as it would show $\mathcal{T}(t) \in \mathcal{T}(R_K)$: we need at most one context switch to fire $a$ from $\mathcal{T}(s')$; the latter visible state belongs to $\mathcal{T}(R_{K-1})$.

So $a$ is a *pop*. Thus, for the thread-visible state $(q, \sigma_i)$ of thread $i$ in state $t$, we have $a = \cdots \to (q, \varepsilon) \in \Delta_i$, for thread $i$'s pushdown program, and after action $a$ the stack of $i$ is either empty ($\sigma_i = \varepsilon$), or symbol $\sigma_i$ was overwritten as part of the action that, some time ago, pushed the symbol just popped, so there is a push edge of the form $\to (?, ?\sigma_i)$, as required by (2). As a result, $\mathcal{T}(t) \in \mathcal{G}$. □

### 4.1.3 Overapproximating the reachable generators

Looking at task (b) below Def. 10, how do we prove that all reachable generators —the elements of $\mathcal{G} \cap \mathcal{T}(R)$—have been reached, for *any* future round? And isn't the computation of $\mathcal{T}(R)$ the problem we set out to solve to begin with, namely the reachability of visible states for arbitrary context bounds? The resolution of the paradox is that we don't need to compute $\mathcal{G} \cap \mathcal{T}(R)$ precisely: any overapproximation of it that is contained in $\mathcal{T}(R_k)$ is sufficient to prove $\mathcal{G} \cap \mathcal{T}(R) \subseteq \mathcal{T}(R_k)$, the crucial condition of Def. 10. An overapproximation of $\mathcal{G} \cap \mathcal{T}(R)$ can in turn be obtained from an overapproximation $Z$ of $\mathcal{T}(R)$; we then have $\mathcal{G} \cap Z \supseteq \mathcal{G} \cap \mathcal{T}(R)$.

How do we statically compute a tight estimate $Z \supseteq \mathcal{T}(R)$? We are overapproximating a set whose exact computation requires a context- and synchronization-sensitive analysis, which is undecidable. However, if we drop **either one** of these sensitivities, the problem becomes decidable and easier in practice. We choose here a *context-insensitive* overapproximation. The tighter this approximation, the weaker the test in Line 4 of Alg. 3, improving the odds for termination.

Our rough idea is that we cut off the stack at size 1. For each push action, we ignore the stack contents underneath the newly pushed element. For each pop action, we don't know what the emerging element is, but we do know that it is either $\varepsilon$ (the stack has become empty), or it is a symbol that was overwritten as part of an earlier push. To cover the latter case, we inspect all push actions in the program and collect the set $E$ of symbols written right underneath the newly pushed symbol. These are the candidates for elements emerging after a pop. We don't care whether it is the right one—our analysis is context-insensitive.

Cutting off the stack at size 1 this way results in a finite-state system that can simply be explored exhaustively; its reachable states will form the set $Z$. Formally, given $\mathcal{P} = (Q, \Sigma, \Delta, q^I)$, we construct $\mathcal{M} = (Q \times \Sigma^{\leq 1}, T)$ as done in Alg. 2. Lines 2–3 collect the set $E$ as defined above. The loop beginning in Line 5 constructs the set $T \subseteq (Q \times \Sigma^{\leq 1}) \times (Q \times \Sigma^{\leq 1})$ of transitions; we write $(q, \sigma) \mapsto (q', \sigma')$ for $((q, \sigma), (q', \sigma')) \in T$. Each action, no matter which type, gives rise to a transition in $\mathcal{M}$ (Line 6; for pushes, the symbol underneath the newly pushed symbol must be dropped, which is accomplished by the $\mathcal{T}$ function). In addition, for pops we context-insensitively overapproximate the emerging symbol, which is done via candidate set $E$, as explained above.

---

**Algorithm 2** Build $\mathcal{M}$ used for computing $Z$

---

**Input**: a sequential pushdown system $\mathcal{P} = (Q, \Sigma, \Delta, q^I)$
**Output**: a sequential finite-state system $\mathcal{M} = (Q \times \Sigma^{\leq 1}, T)$
1: $E = \emptyset$ ▷ set of emerging symbols
2: **for each** push action $(q, \sigma) \to (q', \rho_0 \rho_1) \in \Delta$ **do**
3: $\quad E = E \cup \{\rho_1\}$
4: $T = \emptyset$
5: **for each** action $(q, w) \to (q', w') \in \Delta$ **do**
6: $\quad T = T \cup \{(q, w) \mapsto (q', \mathcal{T}(w'))\}$
7: $\quad$ **if** $w' = \varepsilon$ **then**
8: $\quad\quad$ **for each** $\rho \in E$ **do**
9: $\quad\quad\quad T = T \cup \{(q, w) \mapsto (q', \rho)\}$
10: **return** $\mathcal{M} = (Q \times \Sigma^{\leq 1}, T)$

---

Given a CPDS $\mathcal{P}^n = (\mathcal{P}_1, \ldots, \mathcal{P}_n)$, we build a multithreaded finite state program $\mathcal{M}^n = (\mathcal{M}_1, \ldots, \mathcal{M}_n)$ with $\mathcal{M}_i = (Q \times \Sigma_i^{\leq 1}, T_i)$, $1 \leq i \leq n$, where $\mathcal{M}_i$ is obtained via Alg. 2 on $\mathcal{P}_i$. A *state* of $\mathcal{M}^n$ is an element of $Q \times \Sigma_1^{\leq 1} \times \ldots \times \Sigma_n^{\leq 1}$; a *transition* of $\mathcal{M}^n$, written in the form of $\langle q \mid \sigma_1, \ldots, \sigma_n \rangle \mapsto \langle q' \mid \sigma_1', \ldots, \sigma_n' \rangle$, is defined exactly if there exists $i \in \{1, \ldots, n\}$ such that $(q, \sigma_i) \mapsto (q', \sigma_i')$ and for all $j \neq i, \sigma_j = \sigma_j'$. Like $\mathcal{P}^n$, $\mathcal{M}^n$ is asynchronous: each transition affects the local state of at most one thread. With an initial state $t^I := \langle q^I \mid \varepsilon, \ldots, \varepsilon \rangle$ of $\mathcal{M}^n$, $Z$ is defined as the set of all reachable states of $\mathcal{M}^n$.

**Lemma 12.** $\mathcal{T}(R) \subseteq Z$.

The proof is straightforward and included in App. D. □

**Example 13.** *Fig. 3 shows the two-thread finite-state program $\mathcal{M}^2$ and the set $Z$ for the program in Fig. 1.*

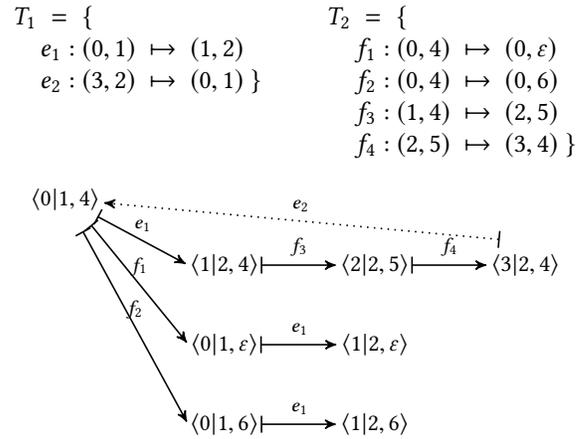

**Figure 3.** An example showing the computation of $Z$ for the CPDS in Fig. 1. $\mathcal{M}^2 = (\mathcal{M}_1, \mathcal{M}_2)$. We start $\mathcal{M}^2$ in $\langle 0 \mid 1, 4 \rangle$. Top: the transitions in $T_1$ and $T_2$; bottom the reachability tree of $\mathcal{M}^2$ whose states form the set $Z$.

### 4.1.4 CUBA via stuttering detection

We summarize the process of verifying property $C$ using $(\mathcal{T}(R_k))_{k=0}^{\infty}$ in Alg. 3 (we assume $R_{-1}$ and $R_0$ are suitably initialized). The algorithm differs from Scheme 1 in the convergence test in Line 4: when reaching a *new* plateau at $k-1$ (no plateau at $k-2$), it launches the generator test. If that fails, we "skip forward" to the next new plateau, if any, since the sets $\mathcal{T}(R_k)$ do not change inside a plateau. This property also makes the algorithm tight, i.e. it stops at the minimal context bound $k$ where $(\mathcal{T}(R_k))_{k=0}^{\infty}$ converges.

The algorithm is partially correct (it does not lie) by the properties of generator sets. It may not terminate if $C$ holds: if $\mathcal{G} \cap Z$ includes unreachable generators, this set will never be contained in $\mathcal{T}(R_k)$, causing the algorithm to run forever.

**Algorithm 3** Verifying $C$ for unbounded $k$ using $(\mathcal{T}(R_k))_{k=0}^{\infty}$ and stuttering detection

1: **for** $k := 1$ **to** $\infty$ **do**
2:    **if** $\mathcal{T}(R_k)$ violates $C$ **then**
3:       **return** "error reachable with resource amount $k$"
4:    **if** $|\mathcal{T}(R_{k-2})| < |\mathcal{T}(R_{k-1})| = |\mathcal{T}(R_k)|$   **and**
      $\mathcal{G} \cap Z \subseteq \mathcal{T}(R_k)$ **then**
5:       **return** "safe for any resource amount"

**Example 14.** *Consider again the program in Fig. 1. Applying Def. 10 we obtain* $\mathcal{G} = \{\langle 0 \mid 1, \varepsilon \rangle, \langle 0 \mid 1, 6 \rangle, \langle 0 \mid 2, \varepsilon \rangle, \langle 0 \mid 2, 6 \rangle\}$. *With set $Z$ as computed in Ex. 13, we obtain* $\mathcal{G} \cap Z = \{\langle 0 \mid 1, \varepsilon \rangle, \langle 0 \mid 1, 6 \rangle\}$. *We now start Alg. 3. The first plateau is reached at $k = 2$, but $\mathcal{G} \cap Z \setminus \mathcal{T}(R_2) = \{\langle 0 \mid 1, 6 \rangle\} \neq \emptyset$. Since visible state $\langle 0 \mid 1, 6 \rangle$ may be reached in the future, we must continue. The next plateau is reached at $k = 5$. Now we indeed have $\mathcal{G} \cap Z \subseteq \mathcal{T}(R_5)$ (visible state $\langle 0 \mid 1, 6 \rangle$ was reached in round 4), so the algorithm terminates; we have $\mathcal{T}(R) = \mathcal{T}(R_5)$.*

To summarize, Alg. 3 gives us a sound method to decide context-unbounded reachability in multi-threaded stack programs. While the observation sequence $(\mathcal{T}(R_k))_{k=0}^{\infty}$ is guaranteed to converge, we had to modify Scheme 1 to account for the possibility of stuttering. In the modification (Alg. 3), we have traded in the convergence guarantee for soundness.

## 5 Finite Context Reachability

Both Scheme 1 ($R_k$) and Alg. 3 are sound but generally non-terminating reachability analyzers for multi-threaded stack programs. There are, however, two differences: first, sequence $(\mathcal{T}(R_k))_{k=0}^{\infty}$ *abstracts* from the reachable state space and can therefore converge faster, as we will demonstrate in Sec. 6. Second, since the sets $\mathcal{T}(R_k)$ are finite, we can always store them using cheaper data structures than the pushdown store automata required for $R_k$, such as extensional containers or BDDs. In particular, no automaton equivalence check is required for fixed point detection. A blemish is that, since we compute $\mathcal{T}(R_k)$ via projection from $R_k$, it seems that we still need to compute $R_k$ in *automaton* form first. Or do we?

It turns out that, in many cases, we do in fact **not**. The insight is that, while the set $R = \cup_{k=1}^{\infty} R_k$ of all reachable global states tends to be infinite in "truly pushdown" programs, *within a single context* only a finite number of new states may be reachable when starting from a finite number of states. By induction, this implies that, for each $k$, set $R_k$ is finite; a condition we call *finite context reachability (FCR)*.

**Example 15.** *Consider the CPDS in Fig. 1. The maximum size of $\mathcal{P}_2$'s call stack after $k$ contexts is roughly $k/2$, rendering the sets $R_k$ finite; this CPDS satisfies FCR. In contrast, the set $R$ of all reachable global states is infinite: the stack grows without bound.*

If all sets $R_k$ are finite, we can represent them using efficient and succinct finite-state data structures. Subsequent uses of $R_k$ then become feasible. For example, the convergence test $R_{k-1} = R_k$ in Scheme 1 ($R_k$) can be done via extensional equality. It is therefore worth investigating how we can decide the FCR condition from the given pushdown program.

To this end we observe: since set $R_k$ is *regular*, it is recognized by a finite-state automaton (FSA). The language of a FSA is finite exactly if every path from an initial state to an accepting state is simple, so finiteness of the reachability set of a PDS is decidable. The dilemma is of course that we don't *have* the PDS that gives rise to $R_k$: we are missing the initial states set, which is precisely $R_{k-1}$.

The solution is to work with approximations that guarantee finiteness. Our plan is as follows. We first show that, if the set of states reachable in a sequential PDS **from all stacks of size $\leq 1$** is finite—a condition that can be effectively checked, as we will discuss—, then the set of states reachable **from any *one* state with arbitrary stack size** is finite. We then use induction on $k$ to show that, if the above condition holds for all threads' PDSs, all sets $R_k$ are finite.

To realize this plan, we need a small amount of additional notation. Let $\mathcal{P} = (Q, \Sigma, \Delta, ?)$ be a PDS (the initial shared state is irrelevant here). For a set of states $S \subseteq Q \times \Sigma^*$, let $R(S)$ denote the set of states reachable in $\mathcal{P}$ starting from $S$ (so $S \subseteq R(S) \subseteq Q \times \Sigma^*$). For a state $s \in Q \times \Sigma^*$ we also write $R(s)$ short for $R(\{s\})$. Finally, we denote by $|s|$ the size of $s$'s stack.

**Lemma 16.** *If $R(Q \times \Sigma^{\leq 1})$ is finite, then for every $s \in Q \times \Sigma^*$, $R(s)$ is finite.*

**Proof**: by induction on $|s|$. For $|s| \leq 1$, $s \in Q \times \Sigma^{\leq 1}$, so $R(s)$ is finite since $R(Q \times \Sigma^{\leq 1})$ is finite. For an arbitrary $s \in Q \times \Sigma^*$ with $|s| \geq 2$, we first show that any path $p$ starting in $s$ can only reach finitely many distinct reachable states (i.e., $p$ is itself finite, or from some state onward along $p$ there are only repeating states). To this end, we distinguish two cases:

(a) If there exists a state $t$ along $p$ with $|t| < |s|$, then, by the induction hypothesis, the suffix of $p$ from $t$ can only reach finitely many distinct reachable states. Since the prefix of $p$ up to $t$ is finite, it also reaches only finitely many reachable states.

(b) Otherwise, the stack size along $p$ never decreases below $|s|$. The bottom frame of $s$'s stack thus never moves to the top (since $|s| \geq 2$, the bottom and top frames are different). The bottom frame thus has no impact on the execution of $p$, but is "carried around" in all states along $p$. Formally, let $p'$ be the sequence of states obtained from $p$ after removing the bottom frame in every state; $p'$ is a valid path in $\mathcal{P}$. Moreover, $p'$'s initial state $s'$ satisfies $|s'| = |s| - 1$, so by the induction hypothesis, $p'$ can only reach finitely many distinct reachable states; call this set

$R_{p'}$. The reachable states along $p$, then, are obtained as

$$R_p = \{\langle q | w\sigma \rangle : \langle q | w \rangle \in R_{p'}\}$$

where $\sigma$ is the bottom stack symbol of $s$. $R_{p'}$ is a finite set, hence so is $R_p$.

The fact that each path $p$ from $s$ only reaches finitely many distinct states is not enough: the number of such paths itself is infinite. The final step is to apply (the contrapositive of) *König's Lemma* about infinite simple paths in locally finite graphs [21] to conclude that the reachability graph spanned by node $s$ is finite.[2]   □

We now move on to step 2 of semi-deciding finite context reachability, which leads to our main result in this section.

**Theorem 17.** *If for all $i \in \{1, \ldots, n\}$, $R(Q \times \Sigma_i^{\leq 1})$ is finite, then for every $k$, $R_k$ is finite.*

**Proof**: by induction on $k$. The condition is true for $k = 0$, since $R_0$ is the set of initial states, which is a singleton by the definition of PDSs. Now assume $R_k$ is finite; its elements are states of the form $\langle q | w_1, \ldots, w_n \rangle$. $R_{k+1}$ is obtained by going through these finitely many states and firing each of the finitely many threads on its corresponding thread state until completion, i.e. as the following finite union:

$$R_{k+1} = \bigcup_{\substack{\langle q|w_1, \ldots, w_n\rangle \in R_k \\ i \in \{1, \ldots, n\}}} \left\{ \begin{array}{l} \text{global states reachable by thread } i \\ \text{executing from } (q, w_i) \end{array} \right\}.$$

To show that $R_{k+1}$ is finite, it therefore suffices to show that, for every thread state $(q, w_i)$ occurring in $R_k$ and every $i$, thread $i$ can only reach finitely many states from $(q, w_i)$. This follows from Lem. 16, since $R(Q \times \Sigma_i^{\leq 1})$ is finite and $(q, w_i) \in Q \times \Sigma_i^*$.   □

What remains to be discussed is how we decide if, for all $i \in \{1, \ldots, n\}$, $R(Q \times \Sigma_i^{\leq 1})$ is finite. This can be done exactly: for each $i$, we build the pushdown store automaton $\mathcal{A}_i$ for the PDS of thread $i$ but with the (finite) initial states set $Q \times \Sigma_i^{\leq 1}$. We then check whether every path in $\mathcal{A}_i$ from an initial state to an accepting state is simple. In fact, this is equivalent to checking the absence of loops in the graph structure of $\mathcal{A}_i$: by the PSA construction [38], any loops in $\mathcal{A}_i$ are connected to the initial and final states sets. If the $\mathcal{A}_i$ have no loops, their languages and hence all sets $R(Q \times \Sigma_i^{\leq 1})$ are finite. As a result, the system satisfies finite context reachability.

As an example, for the two-thread program presented in Fig. 1, we build the PSA for the program and start state set $Q \times \Sigma_i^{\leq 1}$ for $i = 1, 2$. The resulting graphs are loop-free, as illustrated by the left two in Fig. 4, confirming that the CPDS satisfies FCR. Note that the stack size across contexts is unbounded. In contrast, the two-thread program presented in Ex. 8 does not satisfy FCR since there exist self-loops in both threads as shown in the right two graphs of Fig. 4.

---

[2] König's Lemma is today often quoted as follows: *A connected, infinite, and locally finite graph has an infinite simple path.* The reachability graph spanned by node $s$ is locally finite (since the pushdown program $\Delta$ is finite) but not connected. It is, however, "one-way connected", in the sense that all its nodes are reachable from $s$. The 1927 paper by König contains a stronger version of the Lemma that covers the "one-way connected" case.

## 6 CUBA in Practice

We have implemented the ideas described here in a verifier called CUBA, which offers three approaches:

1. Scheme 1 with explicit-state encoding (which requires finite context reachability), denoted Scheme 1 ($R_k$);
2. Alg. 3 with explicit-state encoding (which also requires finite context reachability), denoted Alg. 3 ($\mathcal{T}(R_k)$);
3. Alg. 3 with state sets encoded using pushdown store automata (PSA) $S_k$ ("Symbolic"), denoted Alg. 3 ($\mathcal{T}(S_k)$). Computing the finite projections $\mathcal{T}(S_k)$ from the automata is fairly straightforward and described in App. E.

The overall procedure below shows how these three approaches are organized in CUBA. Given a CPDS $\mathcal{P}^n$ and a property $C$, it first determines whether FCR holds. If so, CUBA forks two computational threads: one runs visible state reachability, and the other one runs global state reachability. It returns the answer of whichever terminates first, if any. Otherwise, CUBA runs visible state reachability with PSA.

---
**Input**: a CPDS $\mathcal{P}^n$ and a property $C$
1: **if** $\mathcal{P}^n$ satisfies FCR **then**
2:    Alg. 3 ($\mathcal{T}(R_k)$) ∥ Scheme 1 ($R_k$)    ▷ two threads
3: **else**
4:    Alg. 3 ($\mathcal{T}(S_k)$)

---

**Empirical Evaluation**

*Experimental Goals.* We empirically evaluate the approaches proposed in this paper against the following questions:

**Q1.** Is CUBA effective? do the observation sequences, either $(R_k)_{k=0}^\infty$ or $(\mathcal{T}(R_k))_{k=0}^\infty$, converge/can stuttering be detected? if so, with small context bounds?

**Q2.** Is FCR effective in practice? That is, is it applicable, and if so will it benefit the computation of $(\mathcal{T}(R_k))_{k=0}^\infty$?

**Q3.** Is CUBA competitive against existing tools?

*Experimental Setup.* CUBA takes CPDS as input. For the evaluation we collected a number of nontrivial concurrent programs, originally written in C or Java. All but one of them are recursive. We first converted them into Boolean programs using predicate abstraction, and then translated the latter into CPDS. For each benchmark, we consider a safety property, specified via an assertion in the original program, or a visible state in the corresponding CPDS, resp. In total, the programs comprise roughly 1,500 lines of code, featuring 5 shared and 3 local variables on average, and are organized into three suites:

**1–3:** a Windows NT Bluetooth driver [13, 36]. The driver has two types of threads: *stoppers*, which call a stopping procedure to halt the driver, and *adders*, which call a procedure to perform I/O. Note that the original

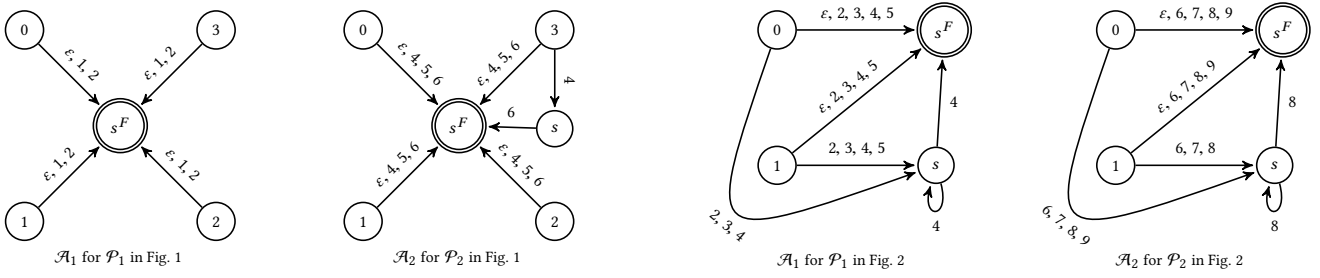

$\mathcal{A}_1$ for $\mathcal{P}_1$ in Fig. 1    $\mathcal{A}_2$ for $\mathcal{P}_2$ in Fig. 1    $\mathcal{A}_1$ for $\mathcal{P}_1$ in Fig. 2    $\mathcal{A}_2$ for $\mathcal{P}_2$ in Fig. 2

**Figure 4.** Determining the FCR condition for the two CPDS in Fig. 1 and 2 via Thm. 17. For $i \in \{1, 2\}$, PSA $\mathcal{A}_i$'s initial states are the shared states of $\mathcal{P}_i$; its accepting state is $s^F$. $\mathcal{A}_i$ accepts $R(Q \times \Sigma_i^{\leq 1})$. The absence of loops in the two automata for Fig. 1 implies their languages and hence $R(Q \times \Sigma_i^{\leq 1})$ are finite.

**Table 2.** Results: *Thread* = # of threads; $n + m^\bullet$ means there are two templates instantiated by $n$ and $m$ threads, resp. Superscript $^\bullet$ indicates the thread is *non*-recursive. *FCR?* = ●: FCR holds; *Safe?* = ✓: assertion holds; $k_{max}$: point of collapse of $(R_k)_{k=0}^\infty$ / $(\mathcal{T}(R_k))_{k=0}^\infty$, ≥ indicates the method was interrupted as the other reached a conclusion. *Time* = runtime (sec); *Mem* = memory usage (MB). Parenthesized number = context bound revealing the bug.

| ID/Program | Prog. Features | | | $(R_k)_{k=0}^\infty$ | $(\mathcal{T}(R_k))_{k=0}^\infty$ | | |
|---|---|---|---|---|---|---|---|
| | Thread | FCR? | Safe? | $k_{max}$ | $k_{max}$ | Time | Mem |
| 1/Bluetooth-1 | 1 + 1 | ● | ✗ | ≥ 7 | 6 (4) | 0.26 | 18.14 |
|  | 1 + 2 | ● | ✗ | ≥ 7 | 6 (3) | 2.32 | 136.26 |
|  | 2 + 1 | ● | ✗ | ≥ 8 | 7 (4) | 12.76 | 347.74 |
| 2/Bluetooth-2 | 1 + 1 | ● | ✗ | ≥ 7 | 6 (4) | 0.53 | 23.43 |
|  | 1 + 2 | ● | ✗ | ≥ 7 | 6 (3) | 4.39 | 196.73 |
|  | 2 + 1 | ● | ✗ | ≥ 8 | 7 (4) | 14.21 | 387.23 |
| 3/Bluetooth-3 | 1 + 1 | ● | ✓ | ≥ 7 | 6 | 0.47 | 22.15 |
|  | 1 + 2 | ● | ✓ | ≥ 7 | 6 | 4.71 | 180.11 |
|  | 2 + 1 | ● | ✓ | ≥ 8 | 7 | 14.46 | 375.42 |

| ID/Program | Prog. Features | | | $(R_k)_{k=0}^\infty$ | $(\mathcal{T}(R_k))_{k=0}^\infty$ | | |
|---|---|---|---|---|---|---|---|
| | Thread | FCR? | Safe? | $k_{max}$ | $k_{max}$ | Time | Mem |
| 4/BST-Insert | 1 + 1 | ● | ✓ | 2 | 2 | 1.17 | 24.53 |
|  | 2 + 1 | ● | ✓ | 3 | 3 | 15.84 | 140.93 |
|  | 2 + 2 | ● | ✓ | ≥ 5 | 4 | 45.21 | 355.74 |
| 5/FileCrawler | $1^\bullet + 2$ | ● | ✓ | 6 | 6 | 0.03 | 5.35 |
| 6/K-Induction | 1 + 1 | ○ | ✓ | ≥ 4 | 3 | 0.23 | 3.78 |
| 7/Proc-2 | $2 + 2^\bullet$ | ○ | ✓ | ≥ 4 | 3 | 0.52 | 18.04 |
| 8/Stefan-1 | 2 | ○ | ✓ | ≥ 3 | 2 | 1.01 | 2.81 |
|  | 4 | ○ | ✓ | ≥ 5 | 4 | 16.36 | 1185.62 |
|  | 8 | ○ | – | ≥ 8 | ≥ 8 | – | OOM |
| 9/Dekker | $2^\bullet$ | ● | ✓ | 6 | 6 | 0.21 | 13.42 |

version of the driver is not recursive; however, we use a recursive procedure to model the counter used in the program, as also done in previous work [13].

**4–5:** 4 implements a binary search tree supporting concurrent manipulations [22]. Two types of threads are considered: an *inserter* inserts nodes to a tree while a *searcher* searches the node with a given value. 5 is an artificial benchmark converted from an online parallel file crawler that allows multiple users to recursively access files in a given directory.

**6–9:** a set of examples taken from previous publications: 6 and 9 from [33], 7 from [13] and 8 from [38].

We conduct two types of experiments: (i) we perform state reachability analysis with on-the-fly assertion checking on each benchmark to empirically answer **Q1** and **Q2**. For unsafe examples, in addition to the context bound that revealed the error, we also report bounds on convergence for all reachable states (which happens later); (ii) we compare the performance of Cuba to that of JMoped [38, 39] to answer **Q3**. As JMoped performs context-bounded analysis, which is unable to prove the correctness of programs, we run it with the same context bound at which Cuba terminates.

All experiments are performed on a 2.3 GHz Intel Xeon machine with 64 GB memory, running 64-bit Linux. The timeout is set to 30 minutes; the memory limit to 4 GB. All benchmarks and our tool are available online [1].

**Results.** Table 2 details our results. Column *FCR* shows that finite context reachability holds in many of our examples.

The various $k_{max}$ columns show the effectiveness of CUBA using global or visible state reachability. We first observe that Alg. 3 terminates on all examples but one, where it runs out of memory[3]. Second, we observe that, in most cases, $k_{max}$ is small, often far less than 10. This is good news as the resource cost increases fast with $k$. One reason is that we compute $(\mathcal{T}(R_k))_{k=0}^\infty$ precisely by projection from $R_k$; the latter's cost is exponential in $k$.

Comparing visible-state to global-state reachability methods, we observe that $(\mathcal{T}(R_k))_{k=0}^\infty$ generally collapses before

---
[3]This example features 8 threads and requires a state set representation using PSAs, which makes the memory usage blow up

$(R_k)_{k=0}^{\infty}$ (it is easy to see that it cannot collapse later). The only way $(R_k)_{k=0}^{\infty}$ can be more effective than $(\mathcal{T}(R_k))_{k=0}^{\infty}$ is that Alg. 3 does not terminate while Scheme 1 $(R_k)$ does, but there is no such case in our benchmarks.

Program #9 is recursion-free (the only one among our benchmarks). Interestingly, although the CUBA fragment over recursion-free programs is decidable, Alg. 3 may still not terminate: we may still not be able to tell stuttering from convergence. On the other hand, Scheme 1 $(R_k)$ is guaranteed to terminate. If the program has a deep call chain, then convergence will not be detected before the stack reaches its maximum depth.

***Tool Comparison.*** Both Cuba and JMoped detected the bugs in Bluetooth suite 1 and 2 (expected), and they did not identify any errors in Bluetooth-3. The running times are comparable, as seen in Fig. 5. The key difference is that, with about the same or less resources, Cuba was able to prove the correctness of Bluetooth-3 and BST-Insert, a huge increase in assurance. As for the implementation: JMoped is built atop the Qadeer/Rehof algorithm [35] and pushdown store automata. Our results show that an explicit-state approach (provided FCR) is competitive and far easier to implement. Although we failed to get to run Getafix, the execution time it reports on the Bluetooth suite [28] is comparable to that of Cuba. However, as with JMoped, Getafix does not prove correctness.

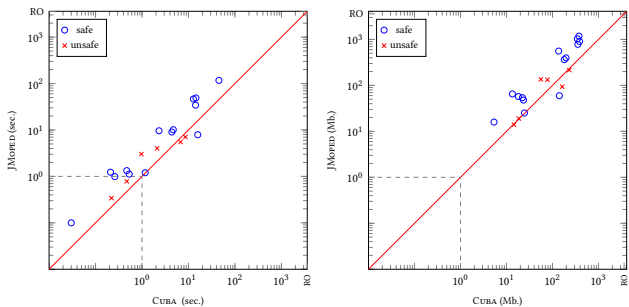

**Figure 5.** Comparing performance of Cuba to JMoped on runtime (left) and memory usage (right). Unsafe: resources used until bug found; safe: until convergence. We run the comparison only on benchmark suites 1–5 and 9, as none of the existing tools, to our knowledge, can properly translate the remaining programs to JMoped format.

## 7 Related Work

Context-sensitive reachability analysis for concurrent recursive programs is undecidable [37], even with only two threads and finite domain variables. In order to cope with undecidability, various remedies have been proposed. One is to limit the computational model so that analysis without a context bound becomes decidable. Restrictions of the threads' synchronization capabilities include no communication between threads [11], communication via a finite number of nested locks [18, 19], communication following a transactional policy [34], or communication via specific concurrent queuing systems [24]. In contrast, our solution applies to the general case of (finite-state) shared-memory concurrency with arbitrary recursion depth. To make progress, we have forgone completeness of the analysis.

Another remedy is to perform context-bounded analysis (CBA), which underapproximates the set of reachable states, by bounding the number of contexts. Originating in the work of Qadeer and Wu [36], CBA has emerged as a practical automatic formal analysis technique for shared-memory concurrent software [9, 10, 24]. The upside is that CBA reduces the concurrent software analysis to a decidable problem [35]. The downside is that a bug which requires more than that bound to manifest will slip through. Our approach is an attempt to eliminate such uncertainty, by comparing the information learned across multiple CBA runs.

Atig et al. [4] suggested a stratified context-bounding method useful for programs with dynamic thread creation. The idea is to bound the number of contexts for each individual thread but not the number of contexts for all threads. However, this essentially limits the size of stack for recursive procedures. In contrast, our work allows each thread to execute with unbounded stack size but the number of threads is fixed. La Torre et al. [26] proposed a partially correct but incomplete strategy that tries to prove all reachable states of a parameterized program are already reached under a $k$-round-robin schedule. Our approach does not enforce such a scheduling to programs.

Inspired by CBA, several fine-grained bounded analyses are proposed and corresponding decidability results are derived. Approaches include: *phase-bounded* analysis [6, 23], where in each phase all pop actions are required to belong to a dedicated thread; *scope-bounded* analysis [27, 40], where the number of scopes (contexts between a procedure call and its return) is bounded; and a method due to [3, 12], which assumes an ordering of the stacks and postulates that a pop operation is subject to the first non-empty stack. Similar to CBA, however, all of them underapproximate.

Prabhu et al. [33] proposed a semi-decision procedure to construct a proof of correctness via combining CBA and $k$-induction. As argued in [33], $k$-induction requires considering paths of length $k$ from arbitrary initial states, which makes the separation from error states non-trivial. Our approach is simpler, as it considers as parameter $k$ the context bound itself and looks for converging sets of reachable states. A potential advantage of [33] is that it investigates inductiveness of an invariant, rather than the whole set of reachable states. If the goal is to prove a safety property that happens to be inductive while the set of reachable states does not converge, then our method will fail while [33] may succeed.

Reducing concurrent to sequential programs and then subjecting them to sequential verifiers is also an active research area [7, 17]. The KISS verifier [36] pioneered this approach via proposing a source-to-source translation of concurrent to sequential programs that underapproximates the behaviors of the original program. It limits the context switches to 2. Lal et al. [29] and La Torre et al. [25] extended the bound from 2 to any fixed $k$. Emmi et al. [15] introduced delay-bounded scheduling and expanded on the capabilities of [29]'s exploration with its ability to analyze programs with dynamic thread creation. A more general sequentialization framework is proposed in [7]. Recent work [31, 32] on sequentialization proposed a translation that allows programs with unbounded contexts. However, it does not allow unbounded recursion. In summary, sequentialization is often more efficient than the Qadeer/Rehof algorithm [35]; several implementations exist. However, it is still under bounded contexts or bounded recursion; it hence retains the limitations of CBA.

Verification of recursive programs is well understood, and tools have been designed and developed. However, most of them focus on sequential programs. JMoped [39] and Getafix [28] can handle concurrent recursive programs, but under context bounds. JMoped implements a BDD-based symbolic version of the Qadeer/Rehof algorithm [35]. Cuba also uses a variant of this algorithm, for programs that do not satisfy finite-context reachability (only for $(\mathcal{T}(R_k))_{k=0}^{\infty}$). However, compared with our tool, the main difference is still that JMoped is built atop context-bounded analysis, and thus mainly a bug-finding tool. Similarly, Getafix is a verifier based on fixed-point calculus and context-bounded analysis.

The technique in this paper can be viewed as solving a parameterized verification problem by determining *cutoffs*. These are bounds on the parameter(s) that provably suffice to draw conclusions about the unbounded program family. In almost all prior works we are aware of, the parameter being cut off is the number of threads or processes [2, 5, 14]. Observation sequences offer a unified approach for arbitrary (discrete) parameters. Also, cutoffs are typically determined statically, often leaving them too large for practical verification. In contrast, our approach is akin to earlier *dynamic* strategies [2, 20]. The work in [20] aims at detecting convergence for thread-count parameters for solving the *decidable* problem of (essentially) local-state reachability in communicating finite-state machines, purely for efficiency.

## 8 Conclusion

In this paper we have applied the paradigm of observation sequences to the context-unbounded analysis of concurrent finite-state recursive procedures. The paradigm resulted in a sound but incomplete method that can both refute and prove safety properties, but may not terminate. Our results support the following conclusions: (i) for our benchmark programs we were able to prove context-unbounded safety in about the same time as, or less time than, previous methods used for context-bounded analysis, and (ii) almost all these programs permit small context bounds not only for error reporting, but also, via sequence convergence, for proving correctness.

A practical open question is whether it may be more efficient to compute the sets $\mathcal{T}(R_k)$ not by projections from the sets $R_k$, but in an abstract interpretation fashion: by defining abstract transfer functions that compute $\mathcal{T}(R_k)$ from $\mathcal{T}(R_{k-1})$. Another practical question is how to improve the scalability of Cuba. There seems to be a dilemma: symbolic representations tend to improve the compactness of state representation, but also make convergence detection more difficult. Finally, an interesting theoretical question is whether the FCR problem is decidable: we have only given sufficient conditions in this work.

## Acknowledgments

We thank the anonymous reviewers and the shepherd for their valuable feedback on this paper. This material is based upon work supported by the U.S. National Science Foundation under Grant No. 1253331. Any opinions, findings, and conclusions or recommendations expressed in this material are those of the authors and do not necessarily reflect the views of the Foundation.

## A  Proofs of Basic Properties

**Property 3.** *If the domain of $OS(O_k)_{k=0}^{\infty}$ is finite, then $(O_k)_{k=0}^{\infty}$ converges.*

**Proof.** Let $K = \{k \in \mathbb{N} : O_k \subsetneq O_{k+1}\}$ be the set of indices where the sequence increases. As the domain of $(O_k)_{k=0}^{\infty}$ is finite, set $K$ is finite. Let therefore $k_0 := \max K + 1$. For every $k \geq k_0$ we have $k \notin K$ and hence, by monotonicity, $O_k = O_{k+1}$. The sequence thus collapses at $k_0$ and converges. □

**Property 4.** *If $OS(O_k)_{k=0}^{\infty}$ does not stutter at $k_0$ and plateaus at $k_0$, then it collapses at $k_0$.*

**Proof.** Recall that "plateaus at $k_0$" means $O_{k_0} = O_{k_0+1}$, and "$(O_k)_{k=0}^{\infty}$ does not stutter at $k_0$" means $O_{k_0} = O_{k_0+1} \Rightarrow \forall k \geq k_0\ O_k = O_{k+1}$ is valid. Together we have $\forall k \geq k_0\ O_k = O_{k+1}$, from which the collapse at $k_0$ immediately follows. □

**Lemma 7.** $(R_k)_{k=0}^{\infty}$ *is stutter-free: for all $k_0$, if $R_{k_0} = R_{k_0+1}$, then for all $k \geq k_0$, $R_k = R_{k+1}$.*

**Proof.** Given $R_{k_0} = R_{k_0+1}$, we show the claim about $k$ by induction. It holds for $k = k_0$. We assume $R_k = R_{k+1}$ and show $R_{k+1} = R_{k+2}$. Appealing to monotonicity, we only show that $R_{k+2} \subseteq R_{k+1}$.

To this end, let $t \in R_{k+2}$, so there exists a path $p$ to $t$ that uses at most $k + 2$ contexts. Let $s$ be the state along this path *right before the final context switch*. State $s$ splits $p$ into two sub-paths: $p = p_1 \circ p_2$.

By construction, $s \in R_{k+1}$. By the assumption $R_k = R_{k+1}$, there exists a path $p'$ to $s$ that uses at most $k$ contexts. Since $p'$ ends in $s$ and $p_2$ starts in $s$, sequence $p'' := p' \circ p_2$ is a well-formed path. Moreover, $p''$ ends in $t$, since $p_2$ does. Finally, $p''$ uses only $k + 1$ contexts: $k$ contexts along $p'$, plus one more along $p_2$. As a result, $t \in R_{k+1}$. □

## B  Concurrent Boolean Procedures

CUBA is intended for concurrent Boolean programs, an abstract finite-state model resulting from predicate abstractions of source code. We present the syntax of a language in the artifact evaluation guide and omitted it here. We present a simplified syntax of a language for such programs in Fig. 6. A formal description of its semantics can be given via a translation to (concurrent) pushdown systems, which is well-studied [28, 35, 38, 39].

$$
\begin{array}{rcl}
prog & ::= & decl^*; func^* \\
decl & ::= & \textbf{decl}\ id^+ \\
func & ::= & type\ id\ (id^*)\ \{\ decl^*;\ [label:\ stmt;]^*\ \} \\
type & ::= & \textbf{void}\ |\ \textbf{bool} \\
stmt & ::= & seqstmt\ |\ \textbf{thread\_create}\ (id) \\
& | & \textbf{atomic}\ \{\ [stmt;]^*\ \}\ |\ \textbf{lock}\ |\ \textbf{unlock} \\
& | & \textbf{while}\ (expr)\ \{\ stmt\ \} \\
& | & \textbf{if}\ (expr)\ \{\ stmt\ \}\ \textbf{else}\ \{\ stmt\ \} \\
seqstmt & ::= & \textbf{skip}\ |\ \textbf{goto}\ label^+ \\
& | & \textbf{assume}\ (expr)\ |\ \textbf{assert}\ (expr) \\
& | & id^+ := expr^+\ [\textbf{constrain}\ expr] \\
& | & id := \textbf{call}\ id\ (expr^*)\ |\ \textbf{call}\ id\ (expr^*) \\
& | & \textbf{return}\ id^*\ <expr> ::= expr\ binop\ expr \\
& | & !expr \\
& | & (expr) \\
& | & const \\
& | & ident \\
& | & * \\
binop & ::= & \text{`\&'}\ |\ \text{`|'}\ |\ \text{`\^{}'}\ |\ \text{`='}\ |\ \text{`!='} \\
const & ::= & \textbf{0}\ |\ \textbf{1}
\end{array}
$$

**Figure 6.** The syntax (partial) of a concurrent language

## C Reachability in PDS

Let $\to^*$ denote the reflexive transitive closure of $\to$. State $s$ of a PDS is *reachable* if $c^I \to^* s$. The reachability problem for PDS is decidable [16, 38]; the idea is as follows. Given a PDS $\mathcal{P} = (Q, \Sigma, \Delta, q^I)$, the set of reachable states of $\mathcal{P}$ can be represented as a standard finite-state automaton $\mathcal{A}$, called *pushdown store automaton* (PSA) [35], defined as $\mathcal{A} = (S, \Sigma, \delta, I, F)$, where

- $S$ is the finite set of states; it satisfies $Q \subseteq S$,
- $\Sigma$ is the input alphabet (same as $\mathcal{P}$'s stack alphabet),
- $\delta \subseteq S \times \Sigma^{\leq 1} \times S$ is the transition relation,
- $I \subseteq S$ is the set of initial states; it satisfies $I \subseteq Q$, and
- $F \subseteq S$ is the set of accepting states; it satisfies $F \cap Q = \emptyset$.

Instead of accepting words, the pushdown store automaton *accepts PDS state* $\langle q\,|\,w \rangle$ if, starting from PSA state $q$ (note: $Q \subseteq S$), reading word $w$ from left to right leads to a PSA state in $F$. We write $L(\mathcal{A})$ for the set of accepted PDS states. The automaton can be constructed in polynomial time [38]. The relationship between $\mathcal{P}$ and $\mathcal{A}$ is given by the following theorem.

**Theorem 8** ([38]). *A PDS state is reachable in $\mathcal{P}$ exactly if it is accepted by $\mathcal{A}$.*

An example of a PDS and its corresponding store automaton is given in Fig. 7.

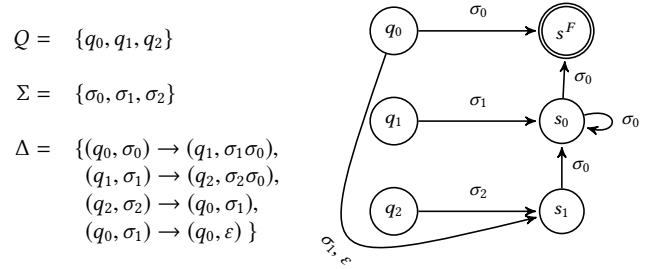

$$
\begin{array}{rcl}
Q & = & \{q_0, q_1, q_2\} \\
\Sigma & = & \{\sigma_0, \sigma_1, \sigma_2\} \\
\Delta & = & \{(q_0, \sigma_0) \to (q_1, \sigma_1 \sigma_0), \\
& & (q_1, \sigma_1) \to (q_2, \sigma_2 \sigma_0), \\
& & (q_2, \sigma_2) \to (q_0, \sigma_1), \\
& & (q_0, \sigma_1) \to (q_0, \varepsilon)\ \}
\end{array}
$$

**Figure 7.** A PDS $\mathcal{P}$ (left) with initial state $\langle q_0\,|\,\sigma_0 \rangle$ and its PSA $\mathcal{A}$ (right), with $I = Q$, $F = \{s^F\}$

## D Proof of Lemma 12

**Lemma 12.** $\mathcal{T}(R) \subseteq Z$.

**Proof**: let $\bar{t} \in \mathcal{T}(R)$; we show $\bar{t} \in Z$. Let $p$ be a path in $\mathcal{P}^n$ to a state $t$ such that $\mathcal{T}(t) = \bar{t}$, and let $\bar{p}$ be the sequence of visible states obtained by applying $\mathcal{T}$ to each state along $p$ in sequence:

$$
\begin{array}{rcccccccccc}
p & :: & t^I = \langle q^I\,|\,\varepsilon, \ldots, \varepsilon \rangle & \xrightarrow{a_1} & \ldots & \xrightarrow{a_{l-1}} & t_{l-1} & \xrightarrow{a_l} & \ldots & \xrightarrow{a_m} & t_m = t \\
& & \downarrow \mathcal{T} & & \downarrow \mathcal{T} & & \downarrow \mathcal{T} & & & & \downarrow \mathcal{T} \\
\bar{p} & :: & \bar{t}^I = \langle q^I\,|\,\varepsilon, \ldots, \varepsilon \rangle & \xmapsto{e_1} & \ldots & \xmapsto{e_{l-1}} & \bar{t}_{l-1} & \xmapsto{e_l} & \ldots & \xmapsto{e_m} & \bar{t}_m = \bar{t}.
\end{array}
$$

We first show that $\bar{p}$ is a legal path in $\mathcal{M}^n$. By the construction of $T$ in Alg. 2, mapping any step $t_l \to t_{l+1}$ along $p$ via $\mathcal{T}$ to an edge $\bar{t}_l \mapsto \bar{t}_{l+1}$ along $\bar{p}$ results in a valid transition in $\mathcal{M}_i$, where $i$ is the index of the thread triggering the step in $p$: all stack symbols but the $i$th are unchanged, since $\mathcal{P}^n$ and $\mathcal{M}^n$ share the same strictly asynchronous execution model.

Since $\bar{p}$ is a legal path in $\mathcal{M}^n$ and ends in $\mathcal{T}(t) = \bar{t}$, visible state $\bar{t}$ is in $\mathcal{M}^n$'s reachability set $Z$. □

---

**Algorithm 4** $\mathcal{T}(\mathcal{A}_i)$

**Input**: A PSA $\mathcal{A}_i = (S_i, \Sigma_i, \delta_i, \{q\}, \{s^F\})$
**Output**: $\{\mathcal{T}(w) \in \Sigma_i^{\leq 1} : (q, w) \in L(\mathcal{A}_i)\}$
1: $E := \emptyset$
2: **for each** pair $(\sigma, q')$ s.t. $(q, \sigma, q') \in \delta_i$ **do**   ▷ label $\sigma$ is a symbol or $\varepsilon$
3:    **if** there is a path from $q'$ to $s^F$ **then**
4:       $E := E \cup \{\sigma\}$
5: **return** $E$

---

## E Implementation of $\mathcal{T}(S_k)$

Alg. 3 ($\mathcal{T}(S_k)$) is identical to Alg. 3 ($\mathcal{T}(R_k)$), except that it uses automata to represent infinite sets of states. It requires the computation of the set $\mathcal{T}(S_k)$ of reachable visible states (which is always finite and represented explicitly) from $S_k$. The latter is a finite set of symbolic states, which take the form $\tau = \langle q \mid \mathcal{A}_1, \ldots, \mathcal{A}_n \rangle$, where $q \in Q$ is a shared state and the $\mathcal{A}_i$ are pushdown store automata. The semantics of symbolic states is given by the concretization function $\gamma$, which maps $\tau$ to a set of concrete states:

$$\gamma(\tau) = \{\langle q \mid w_1, \ldots, w_n \rangle \in Q \times \Sigma_1^* \times \ldots \times \Sigma_n^* : \forall i \, (q, w_i) \in L(\mathcal{A}_i)\} \quad (3)$$

Function $\gamma$ extends to sets $S_k$ of symbolic states pointwise.

Function $\mathcal{T}$ applies to a symbolic state as follows:

$$\mathcal{T}(\langle q \mid \mathcal{A}_1, \ldots, \mathcal{A}_n \rangle) = \{q\} \times \mathcal{T}(\mathcal{A}_1) \times \ldots \times \mathcal{T}(\mathcal{A}_n) \quad (4)$$

where $\mathcal{T}(\mathcal{A}_i) = \{\mathcal{T}(w) \in \Sigma_i^{\leq 1} : (q, w) \in L(\mathcal{A}_i)\}$, $1 \leq i \leq n$. Function $\mathcal{T}$ extends to sets of symbolic states pointwise.

Set $\mathcal{T}(S_k)$ is thus a finite union of subsets of the finite set $Q \times \Sigma_1^{\leq 1} \times \ldots \times \Sigma_n^{\leq 1}$. To compute it, we enumerate the symbolic states $\langle q \mid \mathcal{A}_1, \ldots, \mathcal{A}_n \rangle$ in $S_k$ and apply formula (4). It remains to be described how we determine $\mathcal{T}(\mathcal{A}_i)$, which is the task of Alg. 4.

The **for** loop starting in Line 2 iterates over successors $q'$ of $q$: any stack symbol $\sigma$ appearing at the top of a stack of an accepted state (or $\varepsilon$ if the stack is empty) also appears as label of an edge leaving state $q$ in $\mathcal{A}_i$. Symbol $\sigma$ can be extended to an accepted word on the stack exactly if there is a path from the successor $q'$ to $s^F$, as checked by Alg. 4.